\documentclass[useAMS,usenatbib]{mn2e}

\usepackage{graphicx}

\title[A photometric study of the intermediate polar V515~And]{An extensive photometric study of the recently discovered intermediate polar V515~And (XSS~J00564+4548)}

\author[V. P. Kozhevnikov]{V. P. Kozhevnikov\thanks{E-mail:valerij.kozhevnikov@usu.ru}\\
Astronomical Observatory, Ural Federal University, Lenin Av. 51, Ekaterinburg, 620083, Russia}
 
\begin{document}

\date{Accepted. Received; in original form}

\pagerange{\pageref{firstpage}--\pageref{lastpage}} \pubyear{}

\maketitle

\label{firstpage}

\begin{abstract} 
We report results of photometry of the intermediate polar V515~And. The observations were obtained over 33 nights in 2008 and 2009. The total duration of the observations was 233 h. We clearly detected two oscillations with periods of $465.48493\pm0.00007$ and $488.61822\pm0.00009$~s, which may be the white dwarf spin period and the orbital sideband. The semi-amplitudes of the oscillations are 25 and 20 mmag, accordingly. The oscillation with a period of 465.48493~s has a stable smooth asymmetric pulse profile whereas the pulse profile of the oscillation with a period of 488.61822~s reveals significant changes from a quasi-sinusoidal shape to a shape somewhat resembling a light-curve of an eclipsing binary. Two detected oscillations imply an orbital period of 2.73~h. V515~And is one of the most rapidly spinning intermediate polars with orbital periods less than 3~h and may be not in spin equilibrium. This can be proved by future observations. For this purpose we obtained oscillation ephemerises with a formal shelf life of about 100 yr. (a $1\sigma$ confidence level). 
\end{abstract} 


\section{INTRODUCTION}

Intermediate polars (IPs) form a sub-class of cataclysmic variables (CVs), in which a magnetic white dwarf accretes material from a late type companion filling its Roche lobe. The rotation of the white dwarf is not phase-locked to the binary period of the system. Because the magnetic axis is offset from the spin axis of the white dwarf, this causes oscillations in the X-ray and optical wavelength bands. The X-ray oscillation period is usually identified as the spin period of the white dwarf. In addition to the spin and orbital periods, the reprocessing of X-rays at some part of the system that rotates with the orbital period gives rise to emission that varies with the beat period, where $ 1/P_{\rm beat} = 1/P_{\rm spin} - 1/P_{\rm orb} $. This synodic counterpart is often called the orbital sideband of the spin frequency. A comprehensive review of IPs is given in \citet{patterson94}.

Recently it was shown that IPs possess hard X-ray spectra with high and complex absorption, and this might be sufficient to recognize a CV as an IP \citep{ramsay08}. It was also suggested to call some IPs as "hidden IPs" in the cases when the spin period of the white dwarf is undetectable due to unfavourable alignment of the magnetic dipole axis of the white dwarf \citep{bashkill06, reimer08}. In fact, many X-ray sources detected by {\it Chandra} and {\it INTEGRAL} were classified as IPs on the basis of their hard X-ray spectra \citep{muno04, bird07}. Nonetheless, it is considered that these IP classifications are suggested on the basis of data that cannot cleanly distinguish IPs from other types of CVs and follow-up observations in the optical or softer X-ray bands are needed \citep{pretorius09}. Because the spin period modulation is the defining characteristic of an IP, measurements of the spin period remain the necessary condition for accepting a CV as an IP (Mukai K., http://asd.gsfc.nasa.gov/Koji.Mukai/iphome). Long-term tracking of the spin period is equally important because allows an observational test of spin equilibrium \citep{patterson94}. Moreover, the white dwarf spin rates give insight into the angular momentum flows within the binary \citep{king99}.

The source XSS~J00564+4548 was found in the {\it RXTE} all sky survey and identified with the X-ray {\it ROSAT} source 1RXS J005528.0+461143. Lately it was renamed as V515~And. \citet{bikmaev06} discovered its IP nature. They detected an optical oscillation with a period of about 480~s. \citet{butters08} detected an X-ray period of $465.68\pm0.07$~s and by that confirmed this discovery. Within the measurement errors, this period coincided with the period found by \citeauthor{bikmaev06} and was attributed to the spin period of the white dwarf. In addition, \citeauthor{butters08} found an X-ray oscillation with a period of $489.0\pm0.7$~s, which was interpreted as the orbital sideband. This implies the orbital period must be 2.7~h and places V515~And in the 2--3~h CV period gap. \citet*{bonnet09} found the orbital variability of V515~And both in X-rays and from optical spectra and confirmed this CV in the period gap.  They found a more precise $P_{\rm orb}$ of $2.6244\pm0.0007$~h with possible aliases. However, the oscillation period found in X-rays by \citeauthor{bonnet09} turned out different, namely $469.75\pm0.268$~s. This period is incompatible with $P_{\rm spin}$ found by \citeauthor{butters08} because the difference of the oscillation periods in fifteen times exceeds the measurement error. Although the IP nature of V515~And raises no doubts, obviously $P_{\rm spin}$ cannot reveal such a large instability. To eliminate this uncertainty, in 2008 we performed photometric observations of V515~And. The analysis of these data showed that due to a large oscillation amplitude and a relatively short oscillation period V515~And is a good candidate to search for spin-up or spin-down. To obtain an oscillation ephemeris with a long shelf life necessary for this task, in 2009 we performed additional photometric observations. In this paper we present results of all our observations, spanning a total duration of 233~h within 33 nights.

\section{OBSERVATIONS} \label{observations}

In observations of CVs we use a multi-channel photometer with photomultiplier tubes that allows us to make continuous brightness measurements of two stars and the sky background. Such observations make it possible to obtain evenly spaced data. Then, in cases of smooth signals, classical methods of analysis such as the Fourier transform turn out optimal in comparison with numerous methods appropriated to unevenly spaced data (e.g. Lomb-Scargle periodogram) \citep{schwarzenberg98}. 

V515~And was observed in 2008 October -- December over 15 nights and in 2009 September -- December over 18 nights using the 70-cm telescope at Kourovka observatory, Ural Federal University. A journal of the observations is given in Table~\ref{journal}. The programme and comparison stars were observed through 16-arcsec diaphragms, and the sky background was observed through a 30-arcsec diaphragm. The comparison star is USNO-A2.0 1350-00933560. It has $\alpha=0^h55^m51\fs98$, $\delta=+46\degr14\arcmin4\farcs4$ and $B=14.1$~mag. Data were collected at 8-s sampling times in white light (approximately 300--800~nm), employing a PC-based data-acquisition system. We used the CCD guiding system, which enables precise centring of the two stars in the diaphragms to be maintained automatically. This improves the accuracy of brightness measurements and facilitates the acquisition of long continuous light-curves. The design of the photometer is described in \citet{kozhevnikoviz}.

\begin{table}
\caption{Journal of the observations.}
\label{journal}
\begin{tabular}{@{}l c c}
\hline
\noalign{\smallskip}
Date  &  \hspace{1.5cm} BJD start \hspace{1.5cm} & length \\
(UT) & \hspace{1.5cm} (-245\,4000) \hspace{1.5cm} & (h) \\
\hline
2008 Oct 3   & 743.145179 &  3.6   \\
2008 Oct 4   & 744.144054 & 1.8   \\
2008 Oct 5   & 745.141074 & 9.1   \\
2008 Oct 6   & 746.139669 & 4.6 \\
2008 Oct 7   & 747.140847 & 9.5 \\
2008 Oct 28   & 768.100726 & 10.8 \\
2008 Oct 30     & 770.390577 & 4.1 \\
2008 Oct 31    & 771.399696 & 3.9 \\
2008 Nov 23    & 794.294931& 6.3 \\
2008 Dec 20    & 821.059534 & 11.4 \\
2008 Dec 21   & 822.057118 & 8.6 \\
2008 Dec 22    & 823.057138& 10.6 \\
2008 Dec 23    & 824.057358 & 8.0 \\
2008 Dec 24    & 825.131344  & 8.7 \\
2008 Dec 25    & 826.057337 & 9.0 \\
2009 Sep 14    & 1089.201444 & 2.7 \\
2009 Sep 15   & 1090.183068 & 3.3 \\
2009 Sep 17    & 1092.174982 & 3.4 \\
2009 Sep 26    & 1101.150683 & 9.0 \\
2009 Oct 11    & 1116.138945 & 4.8 \\
2009 Oct 14    & 1119.124073 & 7.6 \\
2009 Oct 22    & 1127.107144 & 7.2 \\
2009 Oct 25   & 1130.109062 & 10.4 \\
2009 Oct 26    & 1131.266053 & 7.0 \\
2009 Nov 13    & 1149.081793 & 8.3 \\
2009 Nov 14    & 1150.310147 & 4.0 \\
2009 Nov 15    & 1151.105588 & 6.8 \\
2009 Nov 16 & 1152.242422 & 8.1 \\
2009 Nov 17   & 1153.072647 & 2.9 \\
2009 Dec 10    & 1176.069215 & 10.2 \\
2009 Dec 15    & 1181.070564& 7.5 \\
2009 Dec 16    & 1182.060185 & 9.4 \\
2009 Dec 17    & 1183.064453 & 7.5 \\

\hline
\end{tabular}
\end{table}

\begin{figure}
\includegraphics[width=84mm]{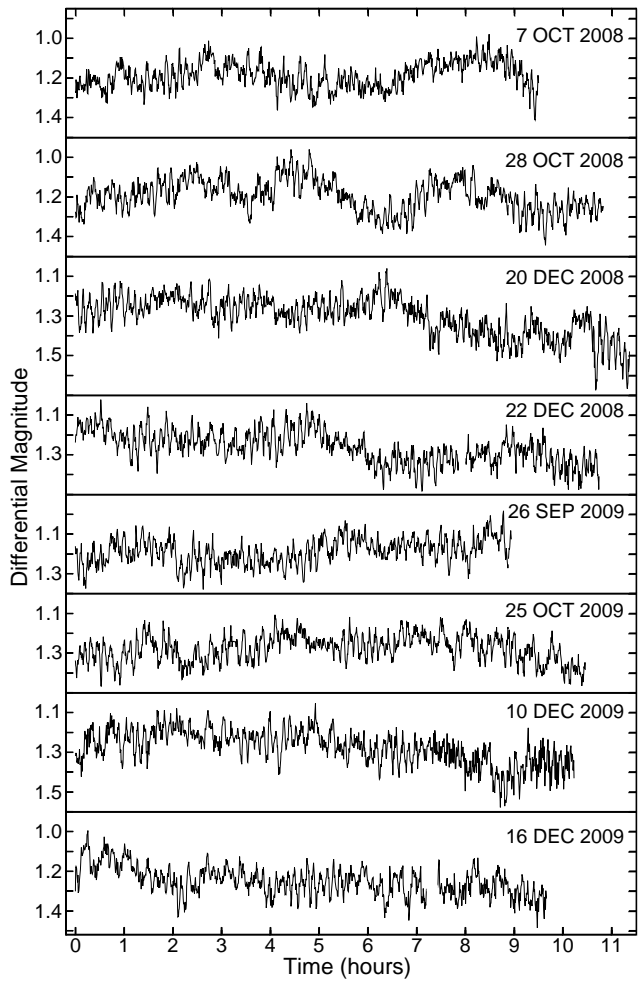}
\caption{Longest differential light-curves of V515~And.}
\label{figure1}
\end{figure}

\begin{figure}
\includegraphics[width=84mm]{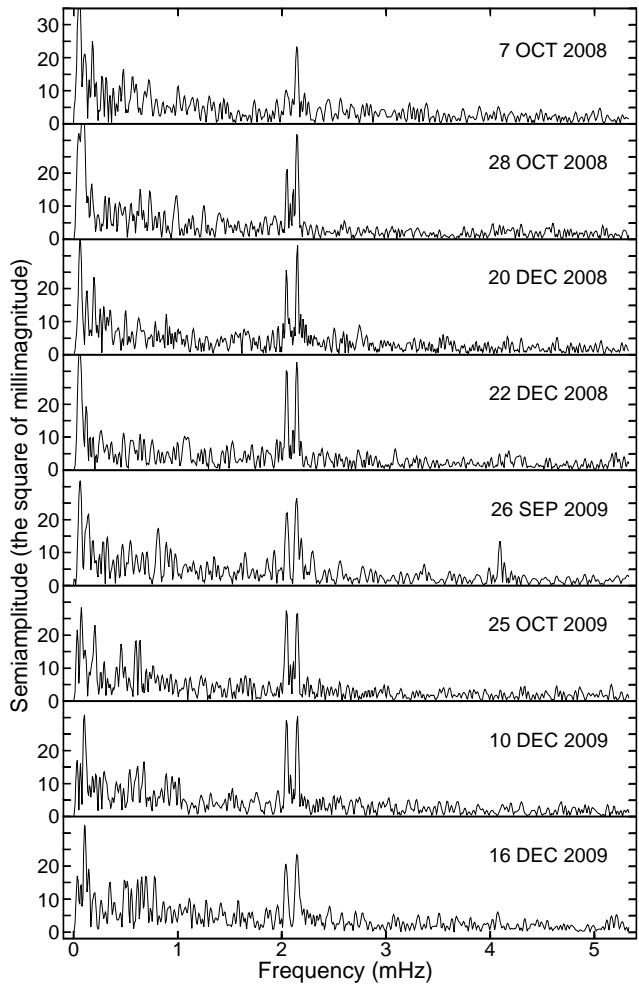}
\caption{Amplitude spectra of V515~And. The prominent peaks visible in the amplitude spectra are appeared with the periods $P_{\rm 1}=466$~s and $P_{\rm 2}=488$~s.}
\label{figure2}
\end{figure}

We obtained differences of magnitudes of the programme and comparison stars taking into account the differences in light sensitivity between the various channels. Because the angular separation between the programme and comparison stars is small, the differential magnitudes were corrected for first order atmospheric extinction and for other unfavorable atmospheric effects (unstable atmospheric transparency, light absorption by thin clouds etc.). According to the mean counts, the photon noise (rms) of the differential light-curves is equal to 28~mmag (a time resolution of 8 s). The actual rms noise also includes atmospheric scintillations and the motion of the star images in the diaphragms. We estimate that these noise components equal approximately 5~mmag each. The total white-noise component of the light-curves (rms) is then 30~mmag. Fig.~\ref{figure1} presents the longest differential light-curves of V515~And, with magnitudes averaged over 32-s time intervals. The white-noise component of these light-curves is 15~mmag.

\section{ANALYSIS AND RESULTS}

As seen in Fig.~\ref{figure1}, the light-curves of V515~And are fairly typical of cataclysmic variables in showing rapid flickering. In addition, periodic oscillations are appreciable in the light-curves.  The light-curves create the impression that these oscillations are unstable because they sometimes disappear. However, this might be due to the interaction of two oscillations with close frequencies.  By using amplitude spectra calculated with the aid of a fast Fourier transform (FFT) algorithm, we revealed that this is exactly the case. These amplitude spectra calculated for longest light-curves are shown in Fig.~\ref{figure2}. Previously low-frequency trends were removed from the light-curves by subtraction of a second-order polynomial fit. Two peaks visible in these power spectra are appeared with the periods $P_{\rm 1}=466$~s and $P_{\rm 2}=488$~s. 

Note that the first harmonic of the oscillation with $P_{\rm 1}$ is absent whereas the first harmonic of the oscillation with $P_{\rm 2}$ sometimes appears.  This is also confirmed by the average power spectrum. Moreover, we found that the first harmonic of $P_{\rm 2}$ appears only in the data of 2009. The average power spectrum of the data of 2008 reveals no high frequency harmonics. On the contrary, the average power spectrum of the data of 2009 reveals the distinct peak at the frequency of the first harmonic of $P_{\rm 2}$, which exceeds in 3.2 times the surrounding noise peaks. But again at the frequency of the first harmonic of $P_{\rm 1}$ this power spectrum does not reveal any prominent peak.  The occurrence of the high frequency harmonic of $P_{\rm 2}$ means that in 2009 the pulse profile of the oscillation with $P_{\rm 2}$ became complicated. Accordingly, the persistent absence of the high frequency harmonic of $P_{\rm 1}$ means that the pulse profile of this oscillation remained smooth and stable.

\begin{figure}
\includegraphics[width=84mm]{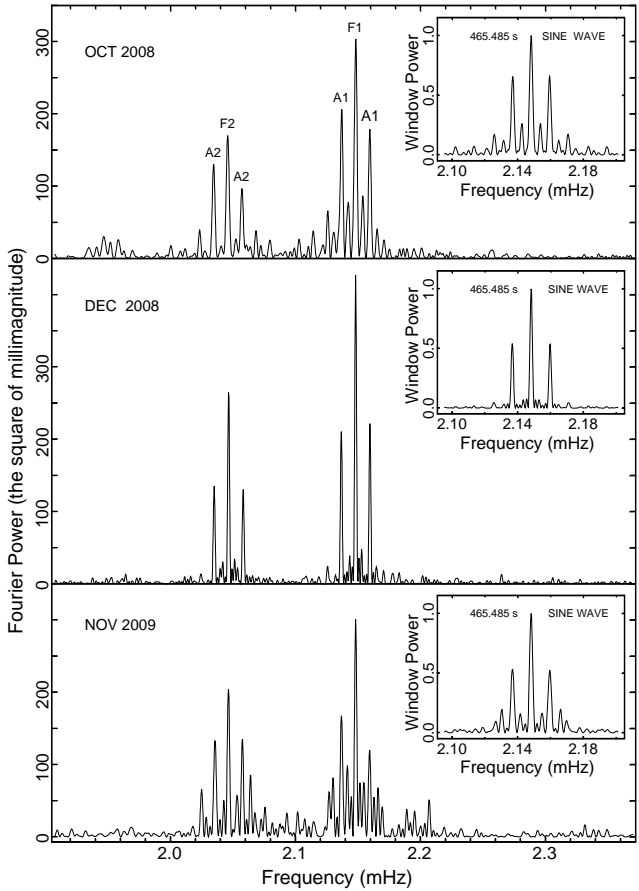}
\caption{Power spectra calculated for three groups of consecutive nights of data from V515~And. They reveal two coherent oscillations with the periods $P_{\rm 1}$ and $P_{\rm 2}$. Inserted frames show the window functions. The principal peaks are labeled with 'F1' and 'F2', and the one-day aliases are labeled with 'A1' and 'A2'.}
\label{figure3}
\end{figure}

A distinctive feature of the periodic oscillations seen in IPs is their coherence because the spin period of the white dwarf is extremely stable. To establish the coherence of oscillations during large time intervals, one can analyse data incorporated into common time series. Then, the coherence can be demonstrated due to coincidence of the structure of the power spectrum in the vicinity of the oscillation frequency and the window function. We begin with three groups of consecutive nights because their window functions are most simple and characterized by the presence of one-day aliases, which directly prove the coherence. Fig.~\ref{figure3} presents the power spectra of the common time series calculated with the FFT algorithm for three groups of data from V515~And consisting of 5 and 6 consecutive nights (see Table~\ref{journal}). The gaps due to daylight and poor weather in these time series were filled with zeroes. To improve the sampling of the power spectra, at the end these time series were supplemented with a considerable number of zeroes.  Previously low frequency trends were removed from the individual light-curves by subtraction of a first or second order polynomial fit. As seen in Fig.~\ref{figure3}, both oscillations reveal distinct pictures closely resembling the window functions obtained from artificial time series consisting of sine waves and the gaps according to the observations. This proves the coherence of the observed oscillations during the corresponding time intervals.

\begin{figure}
\includegraphics[width=84mm]{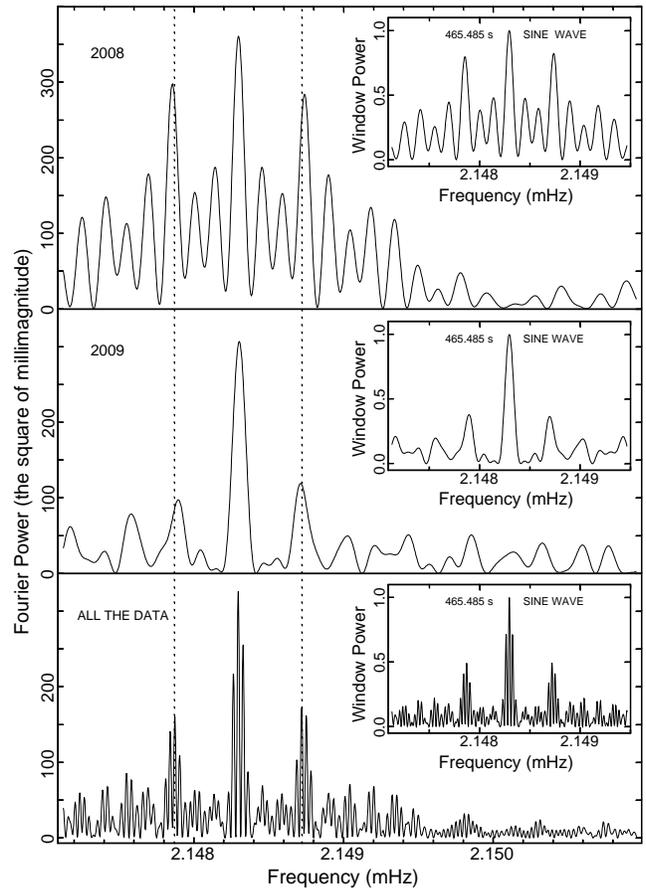}
\caption{Surrounding of the principal peak of the oscillation with the period $P_{\rm 1}$ in the power spectra calculated for the data obtained in 2008 and 2009 and for all the data from V515~And. The dotted lines mark the one-month aliases.}
\label{figure4}
\end{figure}

It is well known that the frequency resolution of data depends on the observational coverage.
Obviously, the highest accuracy of oscillation periods can be achieved from all data incorporated into common time series. However, in such a case the window function may be very complicated due to randomly distributed large gaps, and identification of principal peaks may be difficult. Therefore, we calculated power spectra both for all the data and for the data of 2008 and 2009 separately. These power spectra in the vicinity of $P_{\rm 1}$ are presented in Fig.~\ref{figure4}. The comparison of these power spectra with each other and with the corresponding window functions shows that there are no difficulties in identification of the principal peaks and different aliases and that the oscillation with $P_{\rm 1}$ is coherent during 14 months of observations. The similar picture of the power spectra is also found for the oscillation with $P_{\rm 2}$. 

A half-width of the peak at half maximum (HWHM) in the power spectrum is often accepted as an error of the period because this conforms to the frequency resolution. However, such a method does not allow for the noise of the data and in most cases gives an overestimated error. \citet{schwarzenberg91} showed that the $1\sigma$ confidence interval of the oscillation period is the width of the peak at the $p-N^2$ level, where $p$ is the peak height and $N^2$ is the mean noise power level. Accordingly, the rms error (or $\sigma$) is half of this confidence interval. We used this method to evaluate the precision of the oscillation periods. As the mean noise level we took an average of two levels of the power spectrum in wide frequency intervals prior to and after the oscillation peaks, where the power is unaffected by aliases. The precise maxima of the principal peaks were found by a Gaussian function fitted to the upper parts of the peaks.
 
Having divided our data into six groups, we obtained six measurements of periods for each of the two oscillations. Summarized information is given in Tables~\ref{table2} and \ref{table3}. In addition to the measured periods, in the third columns we give the HWHM of the peaks, which is often used as a conservative error, and in the fourth columns we give the rms errors according to \citet{schwarzenberg91}. One can notice that the rms errors are considerably less than the conservative errors, where this difference is increasing with the length of the observations. Obviously, growth of the precision occurs due to both the increase of the frequency resolution and the decrease of the relative noise level. By using the obtained measurements, we can verify whether the rms error according to \citeauthor{schwarzenberg91} is a real rms error, i.e. whether this error obeys a rule of three sigmas. The errors of the oscillation periods found from all the data are much less than the other errors.  These periods, namely $P_{\rm 1} =465.48493\pm0.00007$~s and $P_{\rm 2}=488.61822\pm0.00009$~s, can be considered absolutely precise with respect to the other periods. Therefore, we can find deviations of the other periods and express them in units of their rms errors. This is shown in the fifth columns of Tables~\ref{table2} and \ref{table3}. In nine cases the deviations are less than the triple rms errors. However, in one case, namely in the data of October 2008 (Table~\ref{table3}), the rms error seems to somewhat depart from the rule of three sigmas because the deviation is $3.4\sigma$.

\begin{table}
\caption{The values and precisions of the period $P_{\rm 1}$.}
\label{table2}
\begin{tabular}{@{}l c c c c c}
\hline
\noalign{\smallskip}
time & period & HWHM & rms & devia- \\
interval & (s) & (s) & error, $\sigma$ (s) & tion \\
\hline
2008 Oct            & 465.469      & 0.27    & 0.031    & $0.5\sigma$  \\ 
2008 Dec           & 465.494      & 0.19    & 0.013     & $0.7\sigma$   \\
2009 Nov           & 465.491      & 0.22    & 0.026     & $0.2\sigma$    \\
2008 all             & 465.4846    & 0.011  & 0.0006   & $0.5\sigma$    \\
2009 all            & 465.4835    & 0.011  &  0.0006  & $2.4\sigma$    \\
total                  & 465.48493 & 0.0018 & 0.00007 & -- \\
\hline
\end{tabular}
\end{table}

\begin{table}
\caption{The values and precisions of the period $P_{\rm 2}$.}
\label{table3}
\begin{tabular}{@{}l c c c c c}
\hline
\noalign{\smallskip}
time & period & HWHM & rms & devia- \\
interval & (s) & (s) & error, $\sigma$ (s) & tion \\
\hline
2008 Oct            & 488.772      & 0.33    & 0.045    & $3.4\sigma$  \\ 
2008 Dec           & 488.598      & 0.20    & 0.019     & $1.1\sigma$   \\
2009 Nov           & 488.634      & 0.26    & 0.037     & $0.4\sigma$    \\
2008 all             & 488.6196    & 0.013  & 0.0008   & $1.7\sigma$    \\
2009 all             & 488.6186    & 0.011  &  0.0007  & $0.5\sigma$    \\
total                   & 488.61822 & 0.0020 & 0.00009 & -- \\
\hline
\end{tabular}
\end{table}

Although this excess of deviation seems small, it is necessary to find plausible reasons for it. This might be caused by the interaction of two oscillations. To find out this, we carried out numerous experiments with artificial time series. These time series were composed of two sine waves with periods and amplitudes, which were similar to periods and amplitudes of the real oscillations, and with the gaps according to the observations. It turned out that the deviations of the periods obtained from the power spectra revealed smooth changes from negative to positive values depending on the initial phase difference between the two sine waves. Because the initial phase difference of the real oscillations is random, we can find total rms errors using the rms errors found from the peak widths and taking into account additions found from the variations of the periods from the artificial time series. On the average, these additions are small and increase the rms errors only by 4 per cent for the oscillation with $P_{\rm 1}$ and by 6 per cent for the oscillation with $P_{\rm 2}$. The maximal addition was found for the oscillation with $P_{\rm 2}$ obtained from the data of October 2008, namely 13 per cent. Obviously, this is due to an unfavourable observational coverage. Taking into account this addition, in this case also the deviation of the period does not exceed $3\sigma$.  

As mentioned, to obtain the high sampling of the power spectra, we added a considerable number of zeroes at the end of the common time series. For the power spectrum of all the data, the sampling, which might allow us to measure the peak width, turned out extremely high, and this power spectrum consisted of $2^{25}$ points. To exclude any gross errors, which might be related with this giant number of points, we also calculated power spectra with the aid of a sine wave fit to the light-curves folded with trial frequencies and found precisely the same oscillation periods. In addition, we used the analysis of variance method \citep{schwarzenberg89}, and this also gave similar results.
 
\begin{figure*} 
\includegraphics[width=176mm]{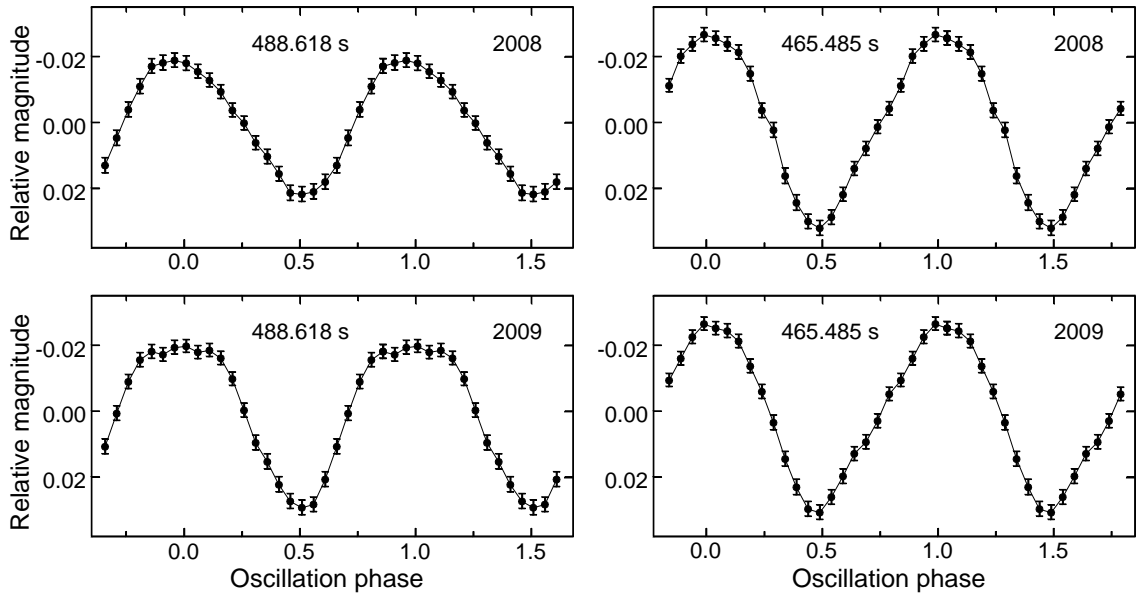}
\caption{Pulse profiles of two oscillations obtained for data of 2008 and 2009 from V515~And. The oscillation with a period of 465.48493~s (on the right) has a stable asymmetric pulse profile whereas the oscillation with a period of 488.61822~s (on the left) reveals a substantially unstable pulse profile.}

\label{figure5}
\end{figure*}

Fig.~\ref{figure5} presents the light-curves of V515~And folded with the periods $P_{\rm 1}=465.48493$~s (on the right) and $P_{\rm 2}=488.61822$~s (on the left)  and obtained by using the data of 2008 and 2009 separately. The oscillation with $P_{\rm 1}$ has a stable smooth asymmetric pulse profile with a slow increase to maximum and with a fast  decline to minimum. It is interesting that the increase is nearly linear whereas the decline is sinuous. The oscillation with $P_{\rm 2}$ reveals a substantially unstable pulse profile, which varies from a quasi-sinusoidal shape to a shape somewhat resembling a light-curve of an eclipsing binary with flat maxima and relatively sharp minima. As follows from the folded light-curves and also from the power spectra, the semi-amplitudes of the oscillations with the periods $P_{\rm 1}$ and $P_{\rm 2}$ are equal approximately to 25 and 20~mmag, accordingly.

The high accuracy of the oscillation periods allows us to obtain oscillation ephemerises with a long shelf life. A rather large noise level in the individual light-curves does not allow us to find oscillation phases directly. Therefore, we found the oscillation phases from the folded light-curves. We decided to obtain the initial phases from the data of 2008 and utilize the data of 2009 for verification. We used 20 phase bins for each folded light-curve; therefore the time interval between adjacent points is 0.05 phases, and is equal to 1/20 of the corresponding period. Obviously, the first points of the folded light-curves are at the distance of 0.025 phases from the very first observational point. Having the initial time of the observations (Table~\ref{journal}) and taking into account that the data were previously averaged over 24~s time intervals, we can find the initial time for the folded light-curves. 

As seen in the folded light-curves (Fig.~\ref{figure5}), the shape of the maximum of the oscillation with $P_{\rm 2}$ is unstable, therefore in the ephemeris of this oscillation we used the time of the minimum. On the contrary, for the oscillation with $P_{\rm 1}$ we used the time of the maximum because its shape is more symmetric. To precisely find these times, we used a Gaussian function fit to 7 points for the minimum and to 9 points for the maximum. Finally we obtained the following ephemerises, in which the initial times of the maximum and minimum were obtained from the observations of 2008:

\begin{equation}
BJD(max)=2454743.146425(23)+0.0053875571(8) E
\end{equation}

\begin{equation}
BJD(min)=2454743.150323(20)+0.0056553035(10) E        
\end{equation}

Using the observations of 2009 and the observations of the three groups of the consecutive nights, we can verify the ephemerises. Table~\ref{table4} gives O -- C values for the maxima and minima obtained from the folded light-curves. As seen, the O -- C values are small and do not exceed a few per cent of the oscillation periods. As the rms errors in table~\ref{table4} we give the summary rms errors of the initial and final maxima or minima without accumulated errors from the oscillation periods. The reason is that we might correct these periods in the case when O -- C values significantly exceed their errors. However, for such a correction we find no reasons. Only in the case of the minimum of 2009 the O -- C value a bit exceeds $3\sigma$, and this deviation is rather caused by the instability of the oscillation profile (see the left part of Fig.~\ref{figure5}).  In addition, in this case the accumulated error of the oscillation period is 5.3~s. Taking into account this error, we find that the O -- C value of the minimum does not exceed $1.5\sigma$. For the maximum of 2009 the accumulated error of the oscillation period is 4.4~s, and the O -- C value does not exceed $0.9\sigma$.

It is considered that a formal shelf life of an ephemeris is the time during that the accumulated error runs up to one oscillation cycle. The accumulated errors from the ephemerises are equal to 4.7~s per year and 5.1~s per year for the oscillations with $P_{\rm 1}$ and $P_{\rm 2}$, accordingly. This corresponds to formal shelf lives of 99 and 84 yr. Here we imply a $1\sigma$ confidence level. Note that the definition of the formal life is rather liberal. For a real shelf life, it seems a reasonable choice to decrease these numbers in 4 times (Mukai~K., http://asd.gsfc.nasa.gov/Koji.Mukai/iphome).

\begin{table}
\caption{Verification of the oscillation ephemerises.}
\label{table4}
\begin{tabular}{@{}l c c c c c}
\hline
\noalign{\smallskip}
time     & total            & O -- C,           & O -- C,          \\
interval & duration (h) & maximum (s)    & minimum (s)   \\
\hline
2008 Oct        & 28.5    & $+5.6\pm4.7$ & $+3.4\pm5.3$  \\ 
2008 Dec       & 56.4    & $-0.7\pm2.2$  & $ 0.0\pm3.1$   \\
2009 Nov       & 30.0    & $+6.7\pm3.1$ & $+1.1\pm3.8$  \\
2009 all          & 123.0  & $+4.3\pm2.2$ & $-8.6\pm2.4$  \\
\hline
\end{tabular}
\end{table}

\begin{figure}
\includegraphics[width=84mm]{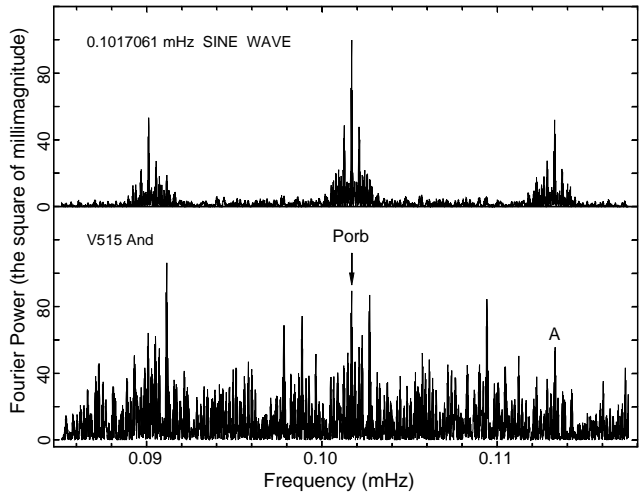}
\caption{Low-frequency part of the power spectrum of the V515~And data, which reveals the peak corresponding to the orbital period.  The one-day alias is also presents (labelled with 'A').}
\label{figure6}
\end{figure}

Two observed oscillations are most probably the white dwarf spin period and orbital sideband. This implies $P_{\rm orb}$ is equal to $2.731086\pm0.000013$~h. In many cases it may be difficult to find the orbital variability of a CV due to rising of the noise level at low frequencies due to flickering. But here this task is somewhat facilitated because we know the precise value of a possible orbital period. Fig.~\ref{figure6} presents the low-frequency part of the power spectrum of the V515~And dada incorporated into the common time series, in which the peak coinciding with the calculated period can be revealed although this peak is not prominent among noise peaks. It corresponds to a period of $2.73118\pm0.00010$~h. The precision of the coincidence of the calculated and detected periods is high, where the difference between them is less than $1\sigma$.

\begin{figure}
\includegraphics[width=84mm]{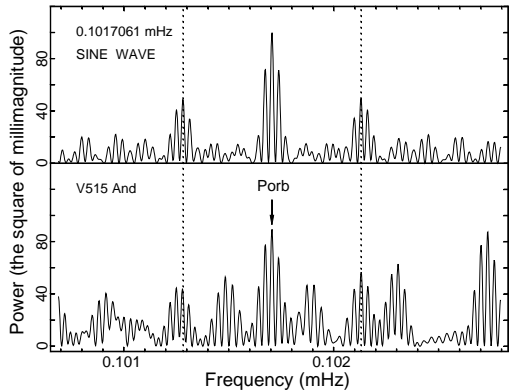}
\caption{Surrounding of the principal peak corresponding to the orbital period. The one-month aliases (denoted by the dotted lines) confirm the reality of the changes with the orbital period.}
\label{figure7}
\end{figure}

\begin{figure}
\includegraphics[width=84mm]{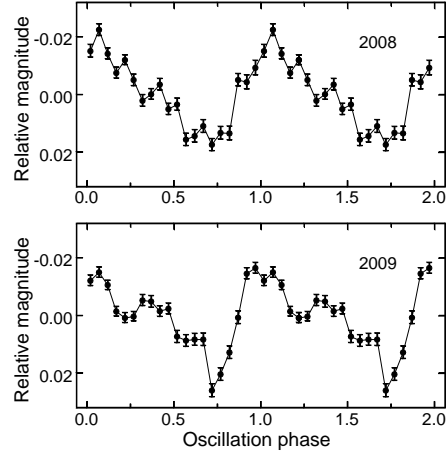}
\caption{ Light-curves of V515~And folded with the orbital period using the data of 2008 and 2009 separately.}
\label{figure8}
\end{figure}

In addition to the coincidence of the calculated and detected periods, we found several signs that the variability with $P_{\rm orb}$ is real. In the power spectrum (Fig.~\ref{figure6}) we revealed the one-day alias, which strictly corresponds to the due frequency. Fig.~\ref{figure7} shows the vicinity of the peak with $P_{\rm orb}$ on expanded scale. Here two one-month aliases are easily detectable. The presence of the one-day and one-month aliases is the sign that the oscillation with $P_{\rm orb}$ is coherent and hence real. Fig.~\ref{figure8} presents two light-curves folded with the detected $P_{\rm orb}$ by using the data of 2008 and 2009 separately. These two folded light-curves show similar shapes and coincide in phase. This again confirms the reality of the orbital variability because not only shows the phase coherence but also reveals the orbital variability within two independent data sets.

\section{DISCUSSION}
We performed extensive photometric observations of V515~And and clearly found two highly coherent oscillations with the periods $P_{\rm 1}=465.48493\pm0.00007$~s and $P_{\rm 2}=488.61822\pm0.00009$~s. We detected these two optical oscillations for the first time. Although the optical oscillation that certifies V515~And as an IP was found earlier by \citet{bikmaev06}, the quality of their observations was such that they did not be able to detect two separate oscillations due to a low frequency resolution. The FWHM of the peak in their Lomb-Scargle plot is $\pm20$~s \citep{butters08}. Obviously, this peak covers both oscillations found by us.

Because the period $P_{\rm 1}$ coincides with the lesser of two periods found by \citeauthor{butters08} in X-rays ($465.68\pm0.07$~s), we can definitely conclude that $P_{\rm 1}$ is the spin period of the white dwarf. We believe that \citeauthor{butters08} give rms errors. Our observations eliminate an ambiguity, which arose from the analysis of X-ray observations made by \citet*{bonnet09}, who found a different spin period, namely $469.75\pm0.268$~s. Obviously, their measurement of the spin period is incorrect.  The period $P_{\rm 2}$ is most probably the orbital sideband. It also coincides with the second period found by \citeauthor{butters08} in X-rays, namely $489.0\pm0.7$~s. \citeauthor{butters08} also interpreted the second period as the orbital sideband. 

The precise knowledge of the spin and sideband periods allows us to calculate the precise value of $P_{\rm orb}$, which is equal to $2.731086\pm0.000013$~h. This $P_{\rm orb}$ agrees with a rough evaluation of $P_{\rm orb}$ made by \citeauthor{butters08} from the spin and sideband periods observed in X-rays. According to \citeauthor{butters08}, $P_{\rm orb}$ must be 2.7~h. Our evaluation of $P_{\rm orb}$ also agrees with $P_{\rm orb}$ evaluated by \citeauthor{bonnet09} from X-ray intensity variations ($2.57\pm0.06$~h) and from radial velocities of two observations taken separately ($2.62\pm0.09$ and $2.82\pm0.21$~h). We also believe that \citeauthor{bonnet09} give rms errors. However, the combined observations of the radial velocities obtained by \citeauthor{bonnet09} give a contradictive value of $P_{\rm orb}$, namely $2.6244\pm0.0007$~h. As \citeauthor{bonnet09} themselves recognize, the combined observations suffer from aliases, and, therefore, this value of $P_{\rm orb}$ may be incorrect. Thus, our observations confirms V515~And is an IP inside the 2--3~h period gap. 

Although the precise coincidence of the peak in the power spectrum with the calculated $P_{\rm orb}$ along with the signs of the coherence of the corresponding oscillation gives evidence of the reality of the orbital variability, this variability is hardly discoverable. This fact allows us to constrain the orbital inclination of the system. The orbital inclination has to be less than about $50^\circ$ because, otherwise, it might be easily detected due to changes of the visibility of the hot spot or eclipses (e.g. \citealt{ladous94}).

V515~And reveals interesting features of the pulse profiles of the detected short-period oscillations. The oscillation with the spin period ($P_{\rm 1}$) shows a stable smooth asymmetric pulse profile, whereas the sideband oscillation ($P_{\rm 2}$) reveals significant changes of the pulse profile from a quasi-sinusoidal shape to a shape somewhat resembling a light-curve of an eclipsing binary. We attempt to find plausible explanations of observed features of the pulse profiles.

In IPs, the optical spin pulses can arise from the hot pole-caps, can occur through reprocessing of X-rays in the axisymmetric parts of the accretion disc and can come from the accretion curtains between the inner disc and the white dwarf \citep{hellier95}. The first possibility can be realized only in a very fast rotator, such as AE~Aqr, in which the size of the accretion curtains is thought to be small and gives a small contribution to the flux compared to the white dwarf \citep{hellier95}. The second possibility can be excluded due to the low inclination of V515~And. The reprocessing requires a sufficient degree of asymmetry, e.g. between the front and the back of the disc. Such asymmetry can occur only in a highly inclined system, which shows eclipses (e.g. DQ~Her, \citealt{petterson80, patterson83}). Thus, as in most other IPs, in V515~And the optical emission with $P_{\rm spin}$  most probably comes from the accretion curtains.

Two-pole disc-fed accretion is believed to be the normal mode of behaviour in IPs. Depending on the sizes and shapes of the accretion curtains both single-peaked and double-peaked spin pulse profiles can be produced. Because these sizes and shapes depend on the magnetic field, in IPs with relatively strong magnetic fields two accreting poles can act in phase so that single-picked, roughly sinusoidal pulse profiles can be produced whereas in IPs with weak magnetic fields two accreting poles can act in anti-phase and produce double-peaked spin pulse profiles \citep{norton99}. It is considered that the rapidly spinning IPs with $P_{\rm spin} <700$~s have weak magnetic fields and therefore usually produce double-peaked pulse profiles (\citeauthor{norton99}). The latter case occurs when the geometry allows the two opposite poles to come into view to the observer (e.g. V405~Aur, \citealt{evans04b}). However, the secondary pole can be continuously hidden by the white dwarf. Then the pulse profile turns out also quasi-sinusoidal. This is seen in NY~Lyp where the emission from one pole only is confirmed by the narrow optical emission lines \citep*{haberl02}. V515~And has $P_{\rm spin}$ much less than 700~s and, in spite of this, shows a quasi-sinusoidal pulse profile. Hence, we must conclude that the secondary pole is continuously hidden by the white dwarf like NY Lyp. The absence of the first harmonic of the oscillation with $P_{\rm spin}$ also suggests that the contribution of the secondary pole acting in anti-phase is negligible.

V515~And clearly reveals significant asymmetry of the quasi-sinusoidal spin pulse profile (the right-hand part of Fig.~\ref{figure5}). This asymmetry seems most difficult for explanation and may be caused by the asymmetry of the accretion curtains. Such an asymmetry can arise due to the interaction of the accretion flow and the magnetic field of the white dwarf when the white dwarf is not in spin equilibrium \citep*{evans04, vrielmann05, evans06}. Thus, in V515~And the asymmetric spin pulse profile can be produced in one asymmetric accretion curtain, where this asymmetry may be caused by non-equilibrium spinning.

In the case of disc-fed accretion the optical orbital sideband arises due to reprocessing of X-rays by structures rotating in the reference frame connected with the orbital motion. Pulse profiles of the sideband oscillation often reveal significant variability. Changes of the structure of the accretion disc are considered to be plausible reasons for such variability. However, these changes are not accompanied by noticeable variability of the star brightness \citep*{woerd84}. In V515~And we also did not find appreciable brightness changes between 2008 and 2009 when a significant change of the pulse profile of the sideband oscillation occurred (the left-hand part of Fig.~\ref{figure5}). In IPs, the accretion disc is much brighter than the secondary star and white dwarf (e.g. \citealt{bonnet01}). Therefore, it seems strange how the accretion disc can change its structure without appreciable brightness changes. 

Another way to produce the orbital sideband is alternation of the accretion flow between two poles of the white dwarf with the sideband frequency. This process occurs in cases of stream-fed and disc-overflow accretion and is the only way to produce the orbital sideband in X-rays \citep{wynn92}. Because the disc-overflow accretion and disc-fed accretion are compatible and can co-exist in an IP simultaneously, the change of the contribution between them can probably produce significant variations of the pulse profile of a sideband oscillation. \citet{butters08} observed the orbital sideband of V515~And in X-rays, and this was the sign of a disc-overflow accretion \citep{hellier93}. But \citet{bonnet09} observed no X-ray sideband. This is the indication that in V515~And the proportion between the disc-fed and disc-overflow accretion can change. Therefore, we can account for the significant change of the pulse profile of the sideband oscillation, which we observed in V515~And, by the change of the contribution between the two accretion modes. It is interesting that a similar change of the pulse profile of the sideband oscillation, namely a variation from a quasi-sinusoidal pulse profile to a pulse profile resembling a light-curve of an eclipsing binary, we observed in MU~Cam (see Fig. 6 in \citealt*{kozhevnikov06}), where the change between the two accretion modes was also observed \citep{staude08}. This change in MU~Cam was accompanied by a large change in optical brightness. Although we observed no appreciable brightness changes of V515~And between 2008 and 2009, the change of the contribution between the two accretion modes in this CV seems yet possible because the disc-overflow accretion can amount to only a small part of the full accretion. Note that a X-ray sideband signal is detected in FO~Aqr, for which a 2 per cent disc overflow fraction has been estimated \citep*{mukai94}. Obviously, the change of the full accretion at a 2 per cent level cannot produce appreciable brightness changes.  

\citet*{norton04} used a model of magnetic accretion to investigate the spin equilibria of magnetic CVs. This allowed them to infer approximate values for the magnetic moments of most known IPs. Many authors used their Fig. 2 to evaluate the magnetic moments of newly discovered IPs (e.g. \citealt{gansicke05, rodriguez05, katajainen10}). Our attempts, however, turned out unsuccessful due to the fact that the $P_{\rm spin}/P_{\rm orb}$ ratio for V515~And, which is equal to 0.047, is found below the lines denoting the spin equilibria for $P_{\rm orb}$ = 3~h or less on Fig. 2 in \citeauthor{norton04} We attempted to find the solution of this problem by changing the mass ratio because the model calculations of \citeauthor{norton04} were made only for the mass ratio $M_{\rm_2}/M_{\rm_1}=0.5$. By using dependence (11) in \citeauthor{norton04} or dependence (3) in \citet{norton08}, we found that a $P_{\rm spin}/P_{\rm orb}$ ratio of 0.047 can intersect the corresponding equilibrium line only if the mass ratio $M_{\rm_2}/M_{\rm_1}$ is larger than 1.5. But such a mass ratio is impossible for an ordinary CV.

Using the Ritter and Kolb catalogue of CVs \citep{ritter03} and the IP home page (Mukai~K., http://asd.gsfc.nasa.gov/Koji.Mukai/iphome), we found 13 possible IPs with $P_{\rm orb} < 3$~h. Among them V515~And is one of the most rapidly spinning IPs. Only WZ~Sge and V455~And have shorter oscillation periods of 27.87 and 67.2~s, accordingly. These two stars, however, show additional oscillations with close periods \citep*{robinson78, araujo05}, and therefore their extremely short periods may be caused by white dwarf pulsations rather than spin of a magnetic white dwarf. In addition, these short periods were not confirmed by X-ray observations. If we exclude WZ~Sge and V455~And from the group of IPs with $P_{\rm orb} < 3$~h then it turns out that V515~And is the most rapidly spinning IP with the shortest $P_{\rm spin}/P_{\rm orb}$ ratio in this group. Obviously, this peculiarity does not allow us to estimate the magnetic moment of the white dwarf by using Fig. 2 in \citet{norton04}. Thus, we can conclude that either V515~And has an unusually low magnetic moment of less than $10^{32}$~G~cm$^3$, which is not embraced by the model calculations, or it substantially deviates from spin equilibrium.

Substantial deviations from spin equilibrium can be detected by measuring of spin-up or spin-down from long-term photometric observations involving an ephemeris with a long shelf life. Our oscillation ephemerises are good for this task. Let us estimate the observational coverage that might result in measurable values of $\dot{P}_{\rm spin}$ for V515~And. From table~1 in \citet{warner96} we learned that for IPs spinning with moderate speed (i.e. excluding AE~Aqr and DQ~Her) the detection threshold of $\dot{P}_{\rm spin}$  is about $10^{-11}$. If we suppose that V515~And possesses such $\dot{P}_{\rm spin}$, then this leads to changes of oscillation phases of about 10~s per year. In Table~\ref{table4} one can see that the random deviations of the oscillation phases from the ephemeris of the oscillation with $P_{\rm 1}$, which has stable pulse profile, do not exceed 7~s when the phases are measured from folded light-curves obtained over a few long nights. Thus, it seems possible to detect spin-up or spin-down in V515~And by observing it over a few long nights per year and by performing these observations during several years.

\section{CONCLUSIONS}

\begin{enumerate}
\item We performed extensive photometric observations of V515~And over 33 nights in 2008 and 2009. 
\item The analysis of these data allowed us to clearly detect two coherent oscillations with the periods $P_{\rm 1}=465.48493\pm0.00007$~s and $P_{\rm 2}=488.61822\pm0.00009$~s, which may be the spin period and orbital sideband.
\item These two oscillations imply the orbital period of the system is equal to $2.731086\pm0.000013$~h. This confirms V515~And is in the period gap. 
\item The oscillation with $P_{\rm 1}$ has a stable smooth asymmetric pulse profile with a nearly linear slow increase to maximum and with a sinuous fast  decline to minimum whereas the pulse profile of the oscillation with $P_{\rm 2}$ reveal significant changes from a quasi-sinusoidal shape to a shape somewhat resembling a light-curve of an eclipsing binary with flat maxima and relatively sharp minima.
\item V515~And is one of the most rapidly spinning intermediate polars with $P_{\rm orb} < 3$~h and may be not in spin equilibrium. 
\item The high accuracy of the oscillation periods, which was achieved due to the large observation coverage and the low noise level in the power spectra, allowed us to obtain the oscillation ephemerises with a formal shelf life of about 100 yr. These ephemerises may be useful for investigations of spin-up or spin-down in V515~And.
\end{enumerate}

\end{document}